# Monolayer phosphorene under time-dependent magnetic field

J. P. G. Nascimento*, V. Aguiar and I. Guedes

Departamento de Física, Universidade Federal do Ceará, Campus do PICI, Caixa Postal 6030, 60455-760, Fortaleza, CE, Brazil.

**Abstract**

We obtain the exact wave function of a monolayer phosphorene under a low-intensity time-dependent magnetic field using the dynamical invariant method. We calculate the quantum-mechanical energy expectation value and the transition probability for a constant and an oscillatory magnetic field. For the former we observe that the Landau level energy varies linearly with the quantum numbers $n$ and $m$ and the magnetic field intensity $B_0$. No transition takes place. For the latter, we observe that the energy oscillates in time, increasing linearly with the Landau level $n$ and $m$ and nonlinearly with the magnetic field. The $(k,l) \to (n,m)$ transitions take place only for $l = m$. We investigate the $(0,0) \to (n,0)$ and $(1,l)$ and $(2,l)$ probability transitions.

*Corresponding author. Tel.: +55 85997797665.

E-mail address: joaopedro@fisica.ufc.br (J. P. G. Nascimento)

## 1. Introduction

Since the production of graphene in 2004 [1] the properties of single and few layers of two-dimensional (2D) materials have attracted great attention owing to possible applications in nanoelectronics. Along the last decade several single layer crystals were also developed as silicene [2], germanene [3], stannene [4], and the transition-metal dichalcogenides [5].

Recently, phosphorene, a single layer of Black Phosphorus [6] has been extensively studied. Due to its high anisotropy, phosphorene exhibits interesting direction-dependent optical and transport properties. Investigations on the energy gap and electronic dispersion of different phosphorene materials can be readily found in Refs. [7-15].

In 2015 Zhou et al [16] investigated theoretically the Landau levels and magneto-transport properties of a monolayer phosphorene under a perpendicular static magnetic field. By considering $\boldsymbol{B} = B_0\boldsymbol{k}$ and the Landau gauge $\boldsymbol{A} = -By\boldsymbol{i}$, they showed that for low-intensity fields the energy of conduction and valence bands varies linearly with the Landau level index $n$ and the magnetic field $B_0$. The Landau splittings of the conduction and valence bands and the respective wavefunctions are different due to the anisotropic effective masses. They also considered the symmetry gauge $\boldsymbol{A} = (A_x, A_y)$ and obtained the wave functions in terms of Laguerre polynomials.

Here we calculate analytically the expressions for the energy and probability transitions of a monolayer phosphorene in the presence of low-intensity time-dependent magnetic field. We use the same Hamiltonian as in Ref. [16] with $\boldsymbol{B} = \boldsymbol{B}(t)$, the symmetry gauge and the Lewis and Riesenfeld method [17] to obtain the analytical solution of the time-dependent Schrödinger equation. We obtain the expressions for the mechanical energy average values and probability transitions between Landau levels. We investigate the cases $\boldsymbol{B}(t) = B_0 \boldsymbol{k}$ and $\boldsymbol{B}(t) = (B_0^2 + B_1^2 cos^2(vt))^{1/2} \boldsymbol{k}$. This paper is outlined as follows. In Section 2, we calculate the time-dependent wave functions of phosphorene. In Section 3, we obtain the mechanical energy average values and probability transitions for the constant and oscillating magnetic field. Finally, the concluding remarks are presented in Section 4.

**2. Eigenfunctions of a single-layer phosphorene in the presence of an external low-intensity magnetic field.**

The time-dependent Schrödinger equation of a single-layer phosphorene in the presence of an external low-intensity magnetic field is given by

$$i\hbar \partial_t |\Psi\rangle = H |\Psi\rangle, \qquad (1)$$

where $|\Psi\rangle = [\psi^c \quad \psi^v]^T$ is a two-component spinor, with $\psi^{c,v}$ corresponding to the envelope functions associated with the probability amplitudes at the respective sublattice sites (the superscript $T$ denotes the transpose of the $[...]$ vector), and

$$H = \begin{pmatrix} E_c + \frac{1}{2}\left(\frac{1}{m'_{cx}}\pi_x^2 + \frac{1}{m_{cy}}\pi_y^2\right) & 0 \\ 0 & E_v - \frac{1}{2}\left(\frac{1}{m'_{vx}}\pi_x^2 + \frac{1}{m_{vy}}\pi_y^2\right) \end{pmatrix}, \qquad (2)$$

is the Hamiltonian of the system, $\boldsymbol{\pi} = (\pi_x, \pi_y) = (p_x - eA_x, p_y - eA_y)$ is the generalized momentum, $e$ is the elementary charge, $\boldsymbol{A} = (A_x, A_y)$ is the magnetic vector potential, $m'_{cx}, m_{cy}, m'_{vx}$ and $m_{vy}$ are effective masses related to the free electron mass $m_e$ ($m'_{cx} = 0.167 m_e, m_{cy} = 0.848 m_e, m'_{vx} = 0.184 m_e$ and $m_{vy} = 1.142 m_e$) and $E_c = 0.34\ eV$ ($E_v = -1.18\ eV$) is the conduction (valence) band edge.

From Eq. (1) we obtain the following uncoupled differential equations

$$i\hbar \partial_t \psi^c(x,y,t) - E_c \psi^c(x,y,t) = \frac{1}{2}\left(\frac{1}{m'_{cx}}\pi_x^2 + \frac{1}{m_{cy}}\pi_y^2\right)\psi^c(x,y,t), \quad (3.a)$$

$$i\hbar \partial_t \psi^v(x,y,t) - E_v \psi^v(x,y,t) = -\frac{1}{2}\left(\frac{1}{m'_{vx}}\pi_x^2 + \frac{1}{m_{vy}}\pi_y^2\right)\psi^v(x,y,t). \quad (3.b)$$

By considering $\boldsymbol{B}(t) = (0,0,B(t))$, choosing the gauge $\boldsymbol{A}(\boldsymbol{r},t) = -(1/2)\boldsymbol{r} \times \boldsymbol{B}(t)$, and redifing the cartesian coordinates as $x = (m_{cy}/m'_{cx})^{1/4} X$ and $y = (m'_{cx}/m_{cy})^{1/4} Y$, Eq. (3.a) reads

$$(i\hbar \partial_t - E_c)\psi^c(X,Y,t) = \left[\frac{1}{2M_c}(P_X^2 + P_Y^2) - \frac{\omega_c(t)}{2}\tilde{L}_z + \frac{M_c \omega_{1c}^2(t)}{2}(X^2+Y^2)\right]\psi^c(X,Y,t), (4)$$

where $P_X = -i\hbar \partial/\partial X$, $P_Y = -i\hbar \partial/\partial Y$, $M_c = \sqrt{m'_{cx} m_{cy}}$, $\omega_c(t) = eB(t)/M_c$, $\tilde{L}_z = XP_Y - YP_X$ and $\omega_{1c}^2(t) = e^2 B^2(t)/4M_c^2$.

By applying the unitary transformation

$$\psi^c(X,Y,t) = U_1 \varphi^c(X,Y,t) = exp\left[i\left(\frac{1}{2\hbar}\tilde{L}_z \int \omega_c(t)dt - \frac{E_c}{\hbar}t\right)\right]\varphi^c(X,Y,t), \quad (5)$$

we can map Eq. (4) into that of a two-dimensional harmonic oscillator with time-dependent frequency, namely

$$i\hbar\partial_t \varphi^c(X,Y,t) = \left[\frac{1}{2M_c}(P_X^2 + P_Y^2) + \frac{M_c \omega_{1c}^2(t)}{2}(X^2 + Y^2)\right]\varphi^c(X,Y,t). \quad (6)$$

Lewis and Riesenfeld showed that an invariant for Eq. (6) is given by [17]

$$I(t) = \frac{1}{2}\left[\left(\frac{X}{\rho_c}\right)^2 + \left(\frac{Y}{\rho_c}\right)^2 + (\rho_c P_X - M_c \dot{\rho}_c X)^2 + (\rho_c P_Y - M_c \dot{\rho}_c Y)^2\right], \quad (7)$$

where $\rho_c(t)$ satisfies the generalized Milne-Pinney [18, 19] equation

$$\ddot{\rho}_c + \omega_{1c}^2(t)\rho_c = \frac{1}{M_c^2 \rho_c^3}, \quad (8)$$

and only real solutions of $\rho_c(t)$ are acceptable to have $I$ hermitian.

According to Lewis and Riesenfeld [17], the invariant $I(t)$ is supposed to satisfy the eigenvalue equation

$$I\phi_{n,m}(X,Y,t) = \lambda_{n,m}\phi_{n,m}(X,Y,t), \qquad (9)$$

where $\lambda_{n,m}$ are time-independent discrete eigenvalues and $\langle \phi_{n,m}, \phi_{n',m'} \rangle = \delta_{nn'}\delta_{mm'}$. The solutions of Eq. (6), $\varphi_{n,m}^c$, are related to the eigenfunctions $\phi_{n,m}$ of $I$ by

$$\varphi_{n,m}^c(X,Y,t) = exp[i\alpha_{n,m}(t)]\phi_{n,m}(X,Y,t), \qquad (10)$$

where the phase functions $\alpha_{n,m}(t)$ satisfy the equation

$$\hbar\frac{d\alpha_{n,m}(t)}{dt} = \langle \phi_{n,m}(X,Y,t) | i\hbar\frac{\partial}{\partial t} - H'(t) | \phi_{n,m}(X,Y,t) \rangle, \qquad (11)$$

and $H'(t)$ corresponds to the right-hand side of Eq. (6).

Next, consider the unitary transformation

$$\phi'_{n,m}(X,Y,t) = U_2\phi_{n,m}(X,Y,t), \qquad (12)$$

where

$$U_2 = exp\left[-\frac{iM_c\dot{\rho}_c}{2\hbar\rho_c}(X^2 + Y^2)\right]. \tag{13}$$

Under this transformation and defining $X = \rho_c r cos\theta$ and $Y = \rho_c r sin\theta$, Eq. (9) now reads

$$I'(t)\sigma_{n,m}(r,\theta) =$$

$$= \left[-\frac{\hbar^2}{2}\left(\frac{\partial^2}{\partial r^2} + \frac{1}{r}\frac{\partial}{\partial r} + \frac{1}{r^2}\frac{\partial^2}{\partial \theta^2}\right) + \frac{r^2}{2}\right]\sigma_{n,m}(r,\theta) = \lambda_{n,m}\sigma_{n,m}(r,\theta), \tag{14}$$

where $I'(t) = U_2 I(t) U_2^\dagger$, $r^2 = \frac{X^2+Y^2}{\rho_c^2}$, $\theta = tan^{-1}\left(\frac{Y}{X}\right)$ and

$$\phi'_{n,m}(X,Y,t) = \frac{1}{\rho_c}\sigma_{n,m}(r,\theta). \tag{15}$$

We decompose $\sigma(r,\theta)$ in the form $\sigma(r,\theta) = R(r)\Theta(\theta)$, where $\Theta(\theta) = e^{im\theta}$, with $m \in \mathbb{Z}$. By defining a new variable $u = \frac{r^2}{\hbar}$ and writing $\sigma(u,\theta) = R(r)\Theta(\theta)$ as

$$\sigma(u,\theta) = (\hbar u)^{\frac{|m|}{2}} e^{-\frac{u}{2}} v(u)\Theta(\theta), \tag{16}$$

Eq. (14) becomes

$$u \frac{\partial^2 v(u)}{\partial u^2} + (|m| + 1 - u) \frac{\partial v(u)}{\partial u} + \frac{1}{2} \left( \frac{\lambda_{n,m}}{\hbar} - |m| - 1 \right) v(u) = 0, \qquad (17)$$

whose solutions are expressed in terms of associated Laguerre polynomial

$$v(u) = L_n^{|m|}(u), \qquad (18)$$

where

$$n = \frac{1}{2} \left( \frac{\lambda_{nm}}{\hbar} - |m| - 1 \right), \qquad (19)$$

$n \in \mathbb{N}$ and $|m| \leq n$. Observe that for each $n$, $m$ can assume $2n + 1$ values.

The normalized eigenfunctions of the invariant $I(t)$ read

$$\phi_{n,m}(X, Y, t) = \left[ \frac{\Gamma(n+1)}{\Gamma(n+|m|+1)} \frac{1}{\pi} \right]^{1/2} \left( \frac{1}{\hbar \rho_c^2} \right)^{\frac{|m|+1}{2}} (X^2 + Y^2)^{\frac{|m|}{2}} e^{im\theta}$$

$$\times \exp\left[ -\frac{1}{2\hbar\rho_c} \left( \frac{1}{\rho_c} - iM_c \dot{\rho}_c \right) (X^2 + Y^2) \right] L_n^{|m|} \left( \frac{X^2 + Y^2}{\hbar \rho_c^2} \right), \qquad (20)$$

and the time-independent eigenvalues are written as $\lambda_{n,m} = \hbar(2n + |m| + 1)$.

The phases $\alpha^c_{n,m}(t)$ which satisfy Eq. (11) are

$$\alpha^c_{n,m}(t) = -(2n + |m| + 1) \int_0^t \frac{dt'}{M_c \rho_c^2(t')}. \qquad (21)$$

Finally, from Eqs. (5), (10), and (20), and returning to original variables, the exact normalized solution of Eq. (3.a) reads

$$\psi^c_{n,m}(r,t) = \left(\frac{1}{\hbar \rho_c^2}\right)^{\frac{|m|+1}{2}} \left[\frac{\Gamma(n+1)}{\Gamma(n+|m|+1)} \frac{1}{\pi}\right]^{\frac{1}{2}} e^{i\alpha^c_{n,m}(t)} \exp\left[i\left(\frac{m}{2}\int \omega_c(t)dt - \frac{E_c}{\hbar}t\right)\right]$$

$$\times \left(\sqrt{\frac{m'_{cx}}{m_{cy}}} x^2 + \sqrt{\frac{m_{cy}}{m'_{cx}}} y^2\right)^{\frac{|m|}{2}} \exp\left[-\frac{1}{2\hbar \rho_c}\left(\frac{1}{\rho_c} - iM_c \dot{\rho}_c\right)\left(\sqrt{\frac{m'_{cx}}{m_{cy}}} x^2 + \sqrt{\frac{m_{cy}}{m'_{cx}}} y^2\right)\right]$$

$$\times e^{im\theta^c_r} L_n^{|m|}\left[\frac{1}{\hbar \rho_c^2}\left(\sqrt{\frac{m'_{cx}}{m_{cy}}} x^2 + \sqrt{\frac{m_{cy}}{m'_{cx}}} y^2\right)\right], \qquad (22)$$

where $\theta^c_r = \tan^{-1}\left[\sqrt{\frac{m_{cy}}{m'_{cx}}} \frac{y}{x}\right]$, $n \in \mathbb{N}$ and $|m| \leq n$.

The solution of Eq. (3.b) is easily obtained by replacing $m'_{cx} \to -m'_{vx}$, $m_{cy} \to -m_{vy}$ and $E_c \to E_v$, namely

$$\psi_{n,m}^v(\boldsymbol{r},t) = \left(\frac{1}{\hbar\rho_v^2}\right)^{\frac{|m|+1}{2}} \left[\frac{\Gamma(n+1)}{\Gamma(n+|m|+1)}\frac{1}{\pi}\right]^{\frac{1}{2}} e^{i\alpha_{n,m}^v(t)} \exp\left[i\left(\frac{m}{2}\int \omega_v(t)dt - \frac{E_v}{\hbar}t\right)\right]$$

$$\times \left(\sqrt{\frac{m'_{vx}}{m_{vy}}}x^2 + \sqrt{\frac{m_{vy}}{m'_{vx}}}y^2\right)^{\frac{|m|}{2}} \exp\left[-\frac{1}{2\hbar\rho_v}\left(\frac{1}{\rho_v} - iM_v\dot\rho_v\right)\left(\sqrt{\frac{m'_{vx}}{m_{vy}}}x^2 + \sqrt{\frac{m_{vy}}{m'_{vx}}}y^2\right)\right]$$

$$\times e^{im\theta_r^v} L_n^{|m|}\left[\frac{1}{\hbar\rho_v^2}\left(\sqrt{\frac{m'_{vx}}{m_{vy}}}x^2 + \sqrt{\frac{m_{vy}}{m'_{vx}}}y^2\right)\right], \tag{23}$$

where $M_v = \sqrt{m'_{vx}m_{vy}}$, $\omega_v(t) = eB(t)/M_v$, $\theta_r^v = \tan^{-1}\left[\sqrt{\frac{m_{vy}}{m'_{vx}}}\frac{y}{x}\right]$ and

$$\alpha_{n,m}^v(t) = -(2n+|m|+1)\int_0^t \frac{dt'}{M_v\rho_v^2(t')}. \tag{24.a}$$

with $\rho_v(t)$ satisfying the Milne-Pinney equation

$$\ddot\rho_v + \omega_{1v}^2(t)\rho_v = \frac{1}{M_v^2\rho_v^3}. \tag{24.b}$$

where $\omega_{1v}^2(t) = e^2B^2(t)/4M_v^2$. For a constant magnetic field Eqs. (22) and (23) agree with those found in Ref. [16].

## 3. Results and discussion.

Using the Cartesian coordinates $x$ as generalized coodinates, the Lagrangian of a particle of mass $m$ and charge $q$ moving in a electromagnetic field reads $L = (m/2)\dot{x}^2 + q\mathbf{A} \cdot \mathbf{x} - q\phi$. The magnetic vector potential $\mathbf{A}$ and the electric scalar potential $\phi$ are in general functions of $\mathbf{x}$ and $t$. According to Ref. [20], the Hamiltonian of this system is the mechanical energy since the "potential" energy in an electromagnetic field is determined only by $\phi$. In this case, the canonical momentum is $\mathbf{p} = m\dot{\mathbf{x}} + q\mathbf{A}$ and the Hamiltonian reads $H = (\mathbf{p} - e\mathbf{A}(\mathbf{r},t))^2/2m + q\phi$. Thus, from Eqs. (22) and (23) we obtain the following quantum mechanical energy expectation values for conduction and valence electrons

$$E^c_{n,m}(t) = E_c + \frac{\hbar}{2(m'_{cx}m_{cy})^{1/2}}\left\{(2n + |m| + 1)\left[\beta_c^2 + \frac{e^2 B^2(t)}{4}\rho_c^2\right] - meB(t)\right\}, (25.a)$$

$$E^v_{n,m}(t) = E_v - \frac{\hbar}{2(m'_{cx}m_{cy})^{1/2}}\left\{(2n + |m| + 1)\left[\beta_v^2 + \frac{e^2 B^2(t)}{4}\rho_v^2\right] - meB(t)\right\}. (25.b)$$

where $\beta_j^2 = \frac{1+\rho_j^2 \dot{\rho}_j^2 M_j^2}{\rho_j^2}$.

Thus, for a given $\mathbf{B}(t)$ one has to solve Eqs. (8) and (24.b) to obtain the mechanical energy $E_{n,m}(t)$ for electrons and holes in a single-layer phosphorene. First, consider the static case $\boldsymbol{B}(t) = B_0 \boldsymbol{k}$. The solutions of Eqs. (8) and (24.b) are $\rho_c = \rho_v = \left(\frac{2}{eB_0}\right)^{1/2}$ and the energy average values for electrons and holes are given by

$$E_{n,m}^c = E_c + \frac{\hbar e B_0}{\left(m'_{cx} m_{cy}\right)^{1/2}} \left(n + \frac{|m| - m + 1}{2}\right), \qquad (26.a)$$

$$E_{n,m}^v = E_v - \frac{\hbar e B_0}{\left(m'_{vx} m_{vy}\right)^{1/2}} \left(n + \frac{|m| - m + 1}{2}\right). \qquad (26.b)$$

The Landau level energy varies linearly with $n$, $m$ and $B_0$. For each $n$, the index $m$ takes $2n + 1$ values. From Eqs. (26) and for a fixed $n$, we have two situations to consider. If $m \geq 0$, we have $n + 1$ degenerated states while, for $m < 0$, $n$ states are nondegenerated.

For an oscillating magnetic field $\boldsymbol{B}(t) = (B_0^2 + B_1^2 \cos^2(vt))^{1/2} \boldsymbol{k}$, Eq. (8) and its solution read, respectively,

$$\ddot{\rho}_c + \frac{e^2}{4M_c^2}(B_0^2 + B_1^2 \cos^2(vt))\rho_c = \frac{1}{M_c^2 \rho_c^3}, \qquad (27.a)$$

$$\rho_c(t) = \left(\frac{1}{|W|M_c}\right)^{\frac{1}{2}} \left\{ MC^2 \left[\frac{e^2 r}{8M_c^2 v^2}, -\frac{e^2 B_1^2}{16M_c^2 v^2}, vt\right] + MS^2 \left[\frac{e^2 r}{8M_c^2 v^2}, -\frac{e^2 B_1^2}{16M_c^2 v^2}, vt\right] \right\}^{\frac{1}{2}}, (27.b)$$

where $r = 2B_0^2 + B_1^2$, $MC$ and $MS$ are the even and odd Mathieu functions, repectively, and $W$ is the Wronskian of $MC$ and $MS$. In Fig. 1 we plot $E_{n,0}^c(t)$ shifted by $-E_c$ for some values of $n$. We observe that $E_{n,0}^c(t)$ oscillates with the same period of the oscillating magnetic field, $T = \pi/v$, and increases with $n$. The oscillation reflects the increasing (decreasing) of the magnetic interaction when the magnetic field increases (decreases).

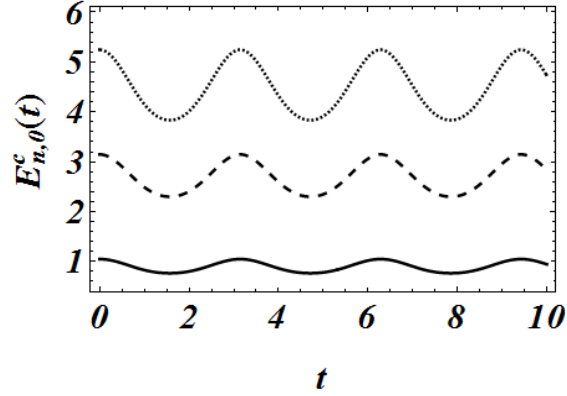

**Fig 1.** Plots of $E_{n,0}^c(t)$ for $n = 0$ (solid line), $n = 1$ (dashed line) and $n = 2$ (dotted line). In this figure we used $\hbar = m_e = v = e = 1$ and $B_0 = B_1 = 0.5$.

In Fig. 2 we plot $E_{0,0}(B_0)$ shifted by $-E_{c,v}$ for three values of $t$ and observe that $E_{0,0}(B_0)$ does not vary linearly with $B_0$. The region defined by the $E_{0,0}(B_0)|_{t=0}$ and $E_{0,0}(B_0)|_{t=\pi/2}$ curves define all possible values of $E_{0,0}(B_0)$, since the period of oscillation is $T = \pi$.

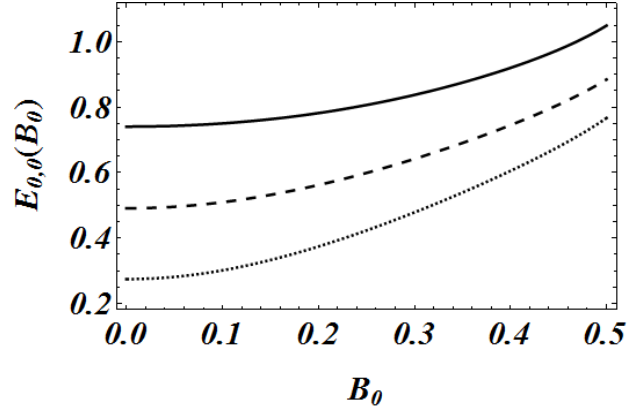

**Fig 2.** Plots of $E_{0,0}(B_0)$ for $t = 0$ (solid line), $t = \pi/4$ (dashed line) and $t = \pi/2$ (dotted line). In this figure we used $\hbar = m_e = \nu = e = 1$ and $B_1 = 0.5$.

The probability of transition from an initial state $\psi_{k,l}^j(\mathbf{r}, t_0)$ to another state $\psi_{n,m}^j(\mathbf{r}, t)$ $(t > t_0)$ reads $P_{(k,l)\to(n,m)}^j(t) = \left|R_{(k,l)\to(n,m)}^j(t)\right|^2$, where $R_{(k,l)\to(n,m)}^j(t)$ is the amplitude transition $\langle n, m, t | k, l, t_0 \rangle$, or

$$R_{(k,l)\to(n,m)}^j(t) = \iint_{-\infty}^{+\infty} dx\,dy \left[\psi_{n,m}^j(\mathbf{r}, t)\right]^* \psi_{k,l}^j(\mathbf{r}, t_0), \qquad (28)$$

where $j$ is $c$ to conduction band electrons and $v$ to valence band holes.

From Eqs. (22) and (23), defining $\rho_j(t) = \rho_j$, $\rho_j(t_0) = \rho_{0,j}$, $\dot{\rho}_j(t) = \dot{\rho}_j$, $\dot{\rho}_j(t_0) = \dot{\rho}_{0,j}$, as well as

$$a_j^2(t) = \frac{1}{\hbar \rho_j^2}, \qquad b_j^2(t_0) = \frac{1}{\hbar \rho_{0,j}^2}, \qquad (29.a)$$

$$B_j(t) = iM_j\left(\rho_j\,\dot\rho_j\,a_j^2(t) - \rho_{0,j}\,\dot\rho_{0,j}\,b_j^2(t_0)\right), \qquad (29.b)$$

$$A^j_{k,l,n,m} = \left(\frac{1}{\hbar\rho_{0,j}^2}\right)^{\frac{|l|+1}{2}} \left(\frac{1}{\hbar\rho_j^2}\right)^{\frac{|m|+1}{2}} \left[\frac{\Gamma(k+1)}{\Gamma(k+|l|+1)}\frac{1}{\pi}\right]^{\frac{1}{2}} \left[\frac{\Gamma(n+1)}{\Gamma(n+|m|+1)}\frac{1}{\pi}\right]^{\frac{1}{2}} e^{-i\alpha^j_{n,m}(t)}$$

$$\times e^{i\alpha^j_{k,l}(t_0)} exp\left[-i\left(\frac{m}{2}\int_0^t \omega_j(t')dt' - \frac{E_j}{\hbar}t\right)\right] exp\left[i\left(\frac{l}{2}\int_0^{t_0}\omega_j(t')dt' - \frac{E_j}{\hbar}t_0\right)\right], (29.c)$$

and using $X = r\cos(\theta_r^j)$ and $Y = r\sin(\theta_r^j)$, where $X = (m_{jy}/m'_{jx})^{-1/4}x$ and $Y = (m'_{jx}/m_{jy})^{-1/4}y$, Eq. (28) reads

$$R^j_{(k,l)\to(n,m)}(t) = \delta_{l,m}\pi A^j_{k,l,n,m}$$

$$\times \int_0^\infty du\, u^{|l|} L_n^{|l|}[a_j^2(t)u]\, L_k^{|l|}[b_j^2(t_0)u]\, exp\left\{-\frac{u}{2}[B_j(t) + a_j^2(t) + b_j^2(t_0)]\right\}, (30)$$

where $u = r^2$ and $L_n^{|l|}(z)$ are the associated Laguerre polynomials. By considering the generating functions

$$\sum_{i=0}^\infty s^i L_i^{|l|}[a_j^2(t)u] = \frac{1}{(1-s)^{|l|+1}} exp\left[-a_j^2(t)u\,\frac{s}{(1-s)}\right], \qquad (|s|<1) \quad (31)$$

and

$$\sum_{\beta=0}^{\infty} w^{\beta} L_{\beta}^{|l|}[b_j^2(t_0)u] = \frac{1}{(1-w)^{|l|+1}} exp\left[-b_j^2(t_0)u\frac{w}{(1-w)}\right], \quad (|w|<1) \quad (32)$$

we found after some algebra

$$R_{(k,l)\to(n,m)}^{j}(t) = \delta_{l,m}\pi A_{k,l,n,m}^{j}$$

$$\times \sum_{p=0}^{k}\left\{\frac{(n+k+|l|-p)!}{p!\,(k-p)!\,(n-p)!}(-1)^p\, 2^{|l|+1}\left[B_j(t) - a_j^2(t) - b_j^2(t_0)\right]^p\right.$$

$$\left.\times \frac{\left[B_j(t) - a_j^2(t) + b_j^2(t_0)\right]^{n-p}\left[B_j(t) + a_j^2(t) - b_j^2(t_0)\right]^{k-p}}{\left[B_j(t) + a_j^2(t) + b_j^2(t_0)\right]^{n+k+|l|+1-p}}\right\}, \quad (33)$$

indicating that the transitions $(k,l) \to (n,m)$ are possible only if $l = m$.

Observe that for a static magnetic field the transition probability is $P_{(k,l)\to(n,m)}^{j}(t) = \delta_{l,m}\delta_{k,n}$ and no transitions are allowed, since $\psi_{k,l}^{j}(\mathbf{r}, t_0)$ and $\psi_{n,m}^{j}(\mathbf{r}, t)$ are stationary states. However, for $\mathbf{B}(t) = (B_0^2 + B_1^2 cos^2(vt))^{1/2}\,\mathbf{k}$ some transitions occur as shown in Figs. 3(a) and (b). From Fig. 3(a) we observe that $P_{(0,0)\to(n,0)}^{c}(t)$ oscillates with $t$ and for $t = pT = \pi p/v$, with $p = 0, 1, 2 ...$, the system will remain in the initial state (0,0). This behavior also reflects the oscillation of the magnetic interaction. The probability for transitions $(0,0) \to (n,0)$ decreases with increasing $n$. In Fig. 3(b) we plot $P_{(1,l)\to(2,l)}^{c}(t)$ for some values of $l$. We observe that $P_{(1,l)\to(2,l)}^{c}(t)$ is symmetric in $l$ and increases with increasing $|l|$.

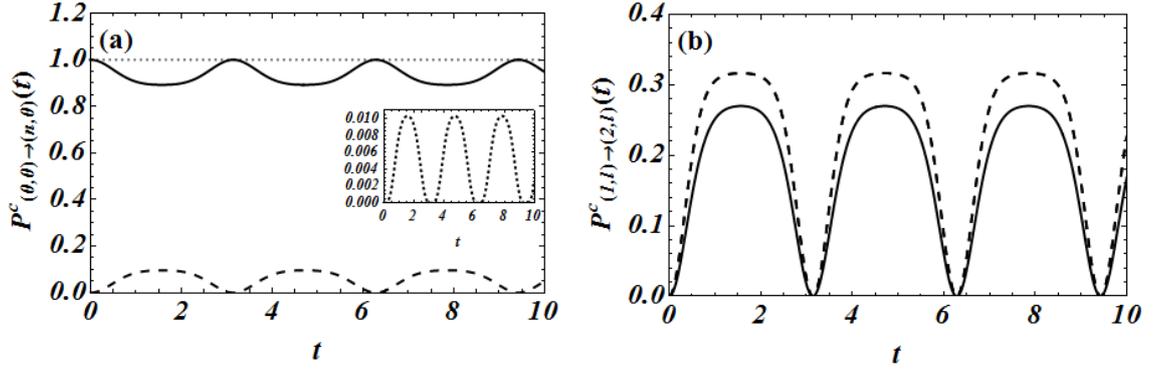

**Fig 3.** Plots of (a) $P^c_{(0,0)\to(n,0)}(t)$ for transitions $n=0$ (solid line), $n=1$ (dashed line) and $n=2$ (dotted line) and (b) $P^c_{(1,l)\to(2,l)}(t)$ for transitions $l=0$ (solid line) and $l=\pm 1$ (dashed line). In this figure we used $\hbar = m_e = \nu = e = 1$ and $B_0 = B_1 = 0.5$.

## 4. Concluding remarks

We obtained the wave function of electrons and holes in a monolayer phosphorene in the presence of a low-intensity time-dependent magnetic field. By considering $\boldsymbol{B} = \boldsymbol{B}(t)$, the symmetry gauge and the Lewis and Riesenfeld method the wave functions found for electrons and holes are given by Eqs. (25.a) and (25.b), respectively. They are expressed in terms of the solutions of the Milne-Pinney equations given by Eqs. (8) and (24.b). From the wave functions obtained we calculated the quantum-mechanical energy expectation value and the transition probability for two magnetic fields, $\boldsymbol{B}(t) = B_0 \boldsymbol{k}$ and $\boldsymbol{B}(t) = (B_0^2 + B_1^2 \cos^2(\nu t))^{1/2}\, \boldsymbol{k}$.

For a constant magnetic field we observe that the quantum-mechanical energy expectation value scales linearly with the Landau levels $n$ and $m$ and the magnetic field $B_0$. In this case, the wave functions are stationary states and no transitions are allowed. For an oscillatory magnetic field we observe that the quantum-mechanical energy expectation value scales linearly with the Landau levels $n$ and $m$ and nonlinearly with the magnetic field. The nonlinear behavior observed with the magnetic field results from the solutions of the Milne-Pinney equations given by the Mathieu functions. The transition probabilities $(0,0) \to (n,0)$ decrease with increasing $n$, while those connecting $(1,l)$ and $(2,l)$ increases with increasing $l$.

**Acknowledgments**

The authors are grateful to the National Counsel of Scientific and Technological Development (CNPq) of Brazil for financial support.